**Computer Science &
Software Engineering**

# Software Requirements Specification of the IUfA's UUIS -- a Team 4 COMP5541-W10 Project Approach

**By**

Abdulrahman Al-Sharawi
Deyvisson Oliveira
Bing Liu
Max Mayantz
Yu Ming Zhang
Ali Alhazmi
Robin de Bled
Kanj Sobh

# Table of Contents

# 1. Introduction

This document presents the business requirement of Unified University Inventory System (UUIS) in Technology-independent manner. All attempts have been made in using mostly business terminology and business language while describing the requirements in this document. Very minimal and commonly understood Technical terminology is used. Use case approach is used in modeling the business requirements in this document.

# 2. Purpose

IUFA purpose is to integrate 3 faculties data bases providing Web interface that allows user to access and manage the integrated inventory.
The IUFA guarantee a secure access to the data from outside university at any time during working hours

# 3. Scope

The IUFA application give the unauthorised user the possibility to use a web based interface that will available to use any time

IUfA involve to the following operations:

- Transferring assets
- Editing assets
- Modifying assets
- Adding inventory assets
- Creating request to borrow an asset or a reserve a space
- Retuning assets
- Creating a new space
- Approving requests
- Authentication
- Search
- Changing permission
- Output reports

# 4. System Description

## 4.1. Inventory assets

Assets in the inventory are classified in 3 types

- Rooms and space
- Software licences
- All other assets

Assets can be grouped like computer parts

## 4.2. University structure

University organizational hierarchy is represented by the following

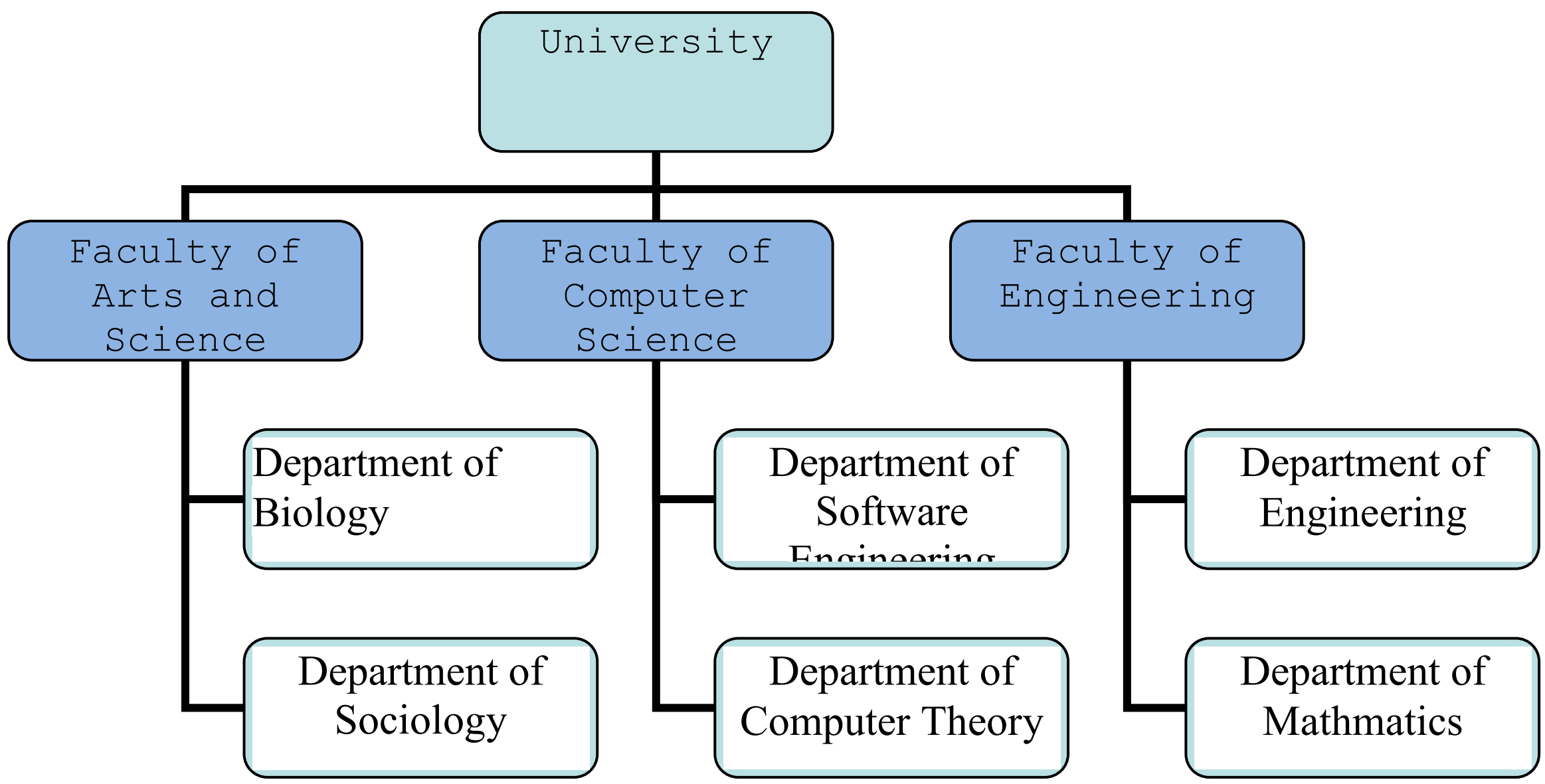

## 4.3. Users levels

Three administrative levels that can approve inventory transactions

- University level (level 3)
- Faculty level (level 2)
- Department level (level 1)
- Users' level can place inventory requests (level 0)
- IT and security level to maintain the inventory (level 4)

## 4.4. User roles

A user can have one of the following role

- University administrators
- Faculty administrators
- Department administrators

- Inventory administrators: users that can be delegated by any administrative level to work and on the applications assigned permission can varies from one user to another depending on its assigned tasks.
  Inventory administrators can be affected to any administrative or organizational level according to their assigned tasks
- Users: students and professors
- IT team system administrators that maintain the system

## 4.5. Permissions List

Permission are should be grouped by role in the following way:

- Department administrator have control on his department inventory
- Faculty administrator have control on his faculty inventory
- University administrator and IT Administrators have the control on the entire inventory
- IT administrators can create different category of permission to help administrators to delegate a part of their tasks

This list resume all permission that a user can have

1. **Request related permissions**
   1.1. request:create
   1.2. request:list
   1.3. request:show
   1.4. request:edit
   1.5. request:aproval or rejection

2. **Asset related permissions**
   **2.1.** asset:create
   **2.2.** asset:list
   **2.3.** asset:show
   **2.4.** asset:edit and modify

3. **Location related permissions**
   **3.1.** location:create
   **3.2.** location:list
   **3.3.** location:show
   **3.4.** location:edit
   **3.5.** location:delete

4. **University related permissions**
   4.1. universityPart:create
   4.2. universityPart:list
   4.3. universityPart:show
   4.4. universityPart:edit

4.5. universityPart:delete

5. **search permission**
   5.1. search:simple
   5.2. search:advanced

6. **reports related permission**
   6.1. report:list
   6.2. report:show

7. **User actions related permission**
   7.1. user:list
   7.2. user:show
   7.3. user:editand change permissions

8. **Audit related permissions**
   8.1. audit:list
   8.2. audit:show

**Note that permissions for each administrative level are predefined**

- **Level 0 has the permission to create requests only**
- **Level 1 can control the assets and locations that he owns**
- **Level 2 can control the assets and locations that owns and get all level1 permissions**
- **Level 3 can control the assets and locations that owns and get all level2 permissions**
- **Level 1, 2 and 3 can delegate some or all of their permission to any user**
- **Level 4 has all possible permissions. Level 4 for can create and assign permission to a new groups**

# 5. Functional requirements

**5.1. Transferring Assets**

5.1.1. Within the same department: data base can be updated directly without any request
5.1.2. Inter departments: request must be approved by a DA group member and faculty group member unless it came from a higher level group
5.1.3. Inter faculties transfer: request can be made by any authorised user and approved by faculty group or higher level
5.1.4. Transfer outside university should be approved by the university group

**5.2. Editing Assets**

5.2.1. Any administrative level user or inventory user can edit an asset that belongs to its department; same thing for faculty user, or university user; in order to make modification if he is authorised to do it.

**5.3. Modifying assets**

5.3.1. all fields of an edited asset can be modified except Ids
5.3.2. a bulk entry file can be used

**5.4. Adding inventory assets**

5.4.1. Any DA group member or authorised inventory group member asset is owned by the department
5.4.2. Any faculty member can add all related departments inventory
5.4.3. Any university group member can add all assets in the inventory
5.4.4. A bulk entry can be used to add many assets

**5.5. Creating request to borrow an asset or a reserve a location**

5.5.1. request can be made by any authorised user
5.5.2. After creation a request still pending waiting to be approved by an administrative level user according to that have this authority

**5.6. Retuning assets**

5.6.1. An inventory user should check returned asset and update inventory

**5.7. Creating a new location**

5.7.1. IT group members can create a new space and modify floor structure when they receive an exception request from any administrative level

**5.8. Approving requests**

5.8.1. Any administration level or authorised inventory group member can display all pending requests waiting for approval from this level and approve those requests
5.8.2. When request is treated user is notified by email
5.8.3. Request is added to the waiting for execution list
5.8.4. Inventory is updated when user receive requested asset

### 5.9. Authentication

5.9.1. Authentication is made by user name and a password for all users

5.9.2. administrative level working on administration computer

### 5.10. Changing permission

5.10.1. Any administrative level user can delegate another user to execute some or all his authorized actions. And this user acquires the role of inventory administrator

### 5.11. Output reports

5.11.1. Asset report by location

5.11.2. Request report

5.11.3. User permission user

# 6. Non functional requirements

## 6.1. Usability

It is mandatory that learning time is between 2 to 4 hours at maximum, because many task are delegated to working students.

Web interface should use clear and consistence terminology in such a way that user; with basic experience on internet and office; find the application easy to use

## 6.2. Availability

The Application should be available always at working hours. Any maintenance or backup operation should be conducted out of working time

## 6.3. Portability

The Application should be installed in any Microsoft or Unix platform

Web application should available to run on browsers like IE, Firefox, Chrome, Opera or Safari

## 6.4. Security

All user are authenticated by user name and a password

Permission are assigned to user according to their roles

Only IT team member can access and maintain data base servers locally

Query is killed if takes more than 1 min

Backup operation are executed periodically

## 6.5. Maintainability

It is important to design system to facilitated future evolution and facilitate maintenance operations

# 7. USE CASES

**Name:** Modify Use Case
**Identifier**: MOD.UC
**Description**
The use case describes the modification that the Inventory Admin can do.
**Goal**
The Inventory Admin initiates the use case. The use case presents all the modification that can be done by the Inventory Admin.
**Preconditions**

1. The Inventory Admin is authenticated

**Assumptions**

1. We assume that use Knows the results of each operation there is no go back actions

**Basic Course**

1. Use case begins when Inventory Admin start searching for an asset
2. Inventory Admin Edit the asset
3. Inventory Admin modify asset properties

**Alternate Course A:**
**Condition:** administrator or authorised inventory user is working on waiting for approval list or waiting for execution list

1. Inventory user or Admin Edit the asset
2. Inventory user or Admin modify asset properties

**Exceptional Course :**

1.
    1. Inventory Admin search for asset
    2. Inventory Admin edit asset
    3. Inventory Admin asset out of inventory
    4. Message error because asset cannot be modified
2.
    1. Inventory Admin search for asset
    2. Inventory Admin edit asset
    3. Inventory Admin does not have sufficient privileges to edit asset
    4. Message error is displayed
3.
    1. Inventory Admin search for asset
    2. no asset found
    3. Message error is displayed

**Post conditions**

1. The system state change according to modification

**Actors**
Inventory Admin, Inventory system, Authentication system

**Included Use Cases**
1. Search use case
2. Edit use case
3. Authentication use case

**Notes**
*We suppose that Hosting system and servers support all the operations*

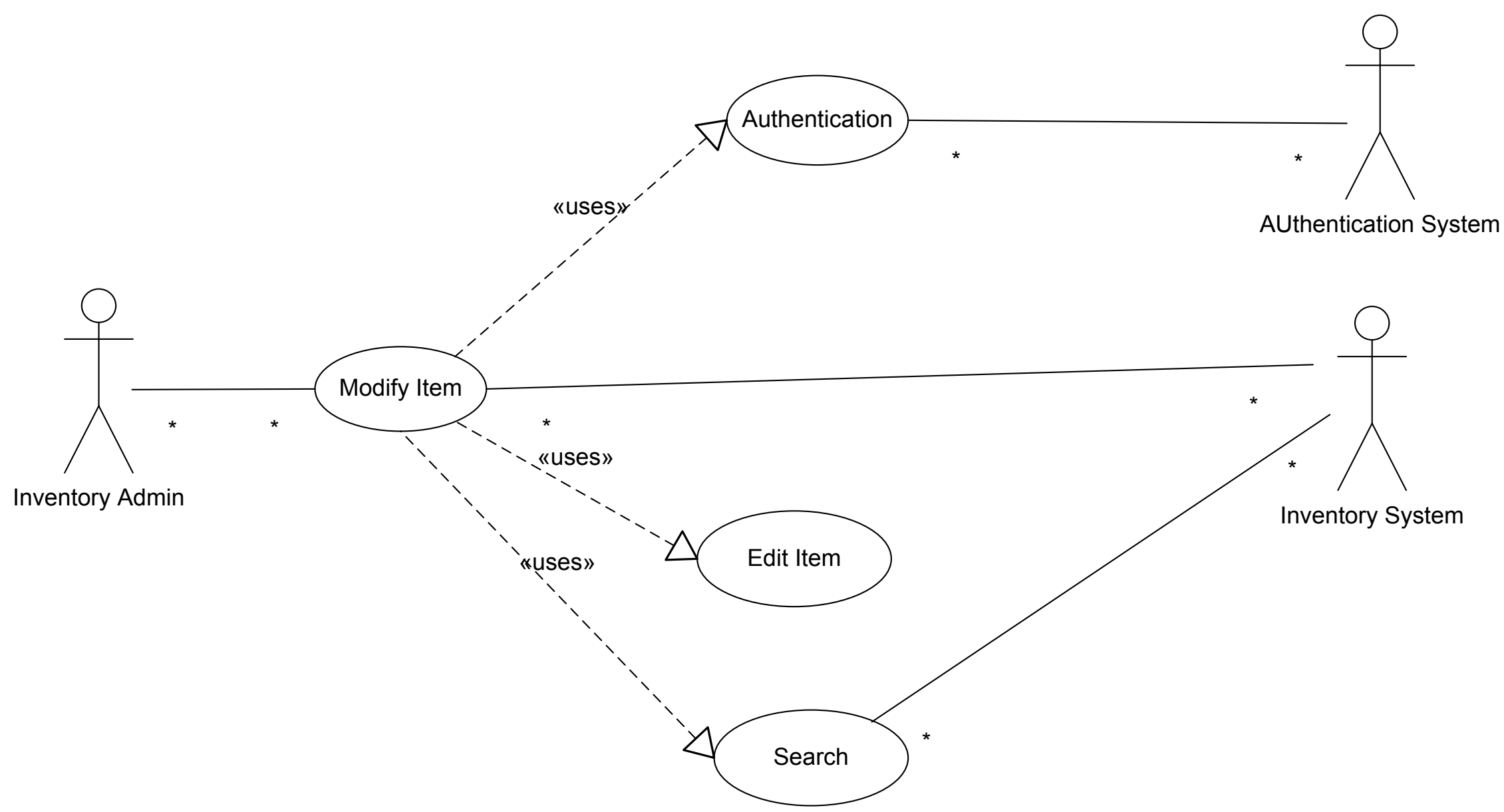

**MOD.UC Modify Use Case**

**Name:** Edit Use Case
**Identifier**: EDT.UC
**Description**
The use case describes the edit operation that the Inventory Admin can do.
**Goal**
The Inventory Admin initiates the use case. The use case presents the edit done by the Inventory Admin.
**Preconditions**

1. The Inventory Admin is authenticated

**Assumptions**

1. We assume that use Knows the results of each operation there is no go back actions

**Basic Course**

1. Use case begins when Inventory Admin start searching for an asset
2. Inventory Admin Edit the asset

**Alternate Course A:**
**Condition:** administrator or authorised inventory user is working on waiting for approval list or waiting for execution list

1. Inventory user or Admin Edit the asset

**Exceptional Course :**

1.
    1. Inventory Admin search for asset
    2. Inventory Admin edit asset
    3. Inventory Admin does not have sufficient privileges to edit asset
    4. Message error is displayed
2.
    1. Inventory Admin search for asset
    2. no asset found
    3. Message error is displayed

**Post conditions**

1. The system state change according to modification

**Actors**
Inventory Admin, Inventory system, Authentication system

**Included Use Cases**

1. Search use case
2. Authentication use case

**Notes**

*We suppose that Hosting system and servers support all the operations*

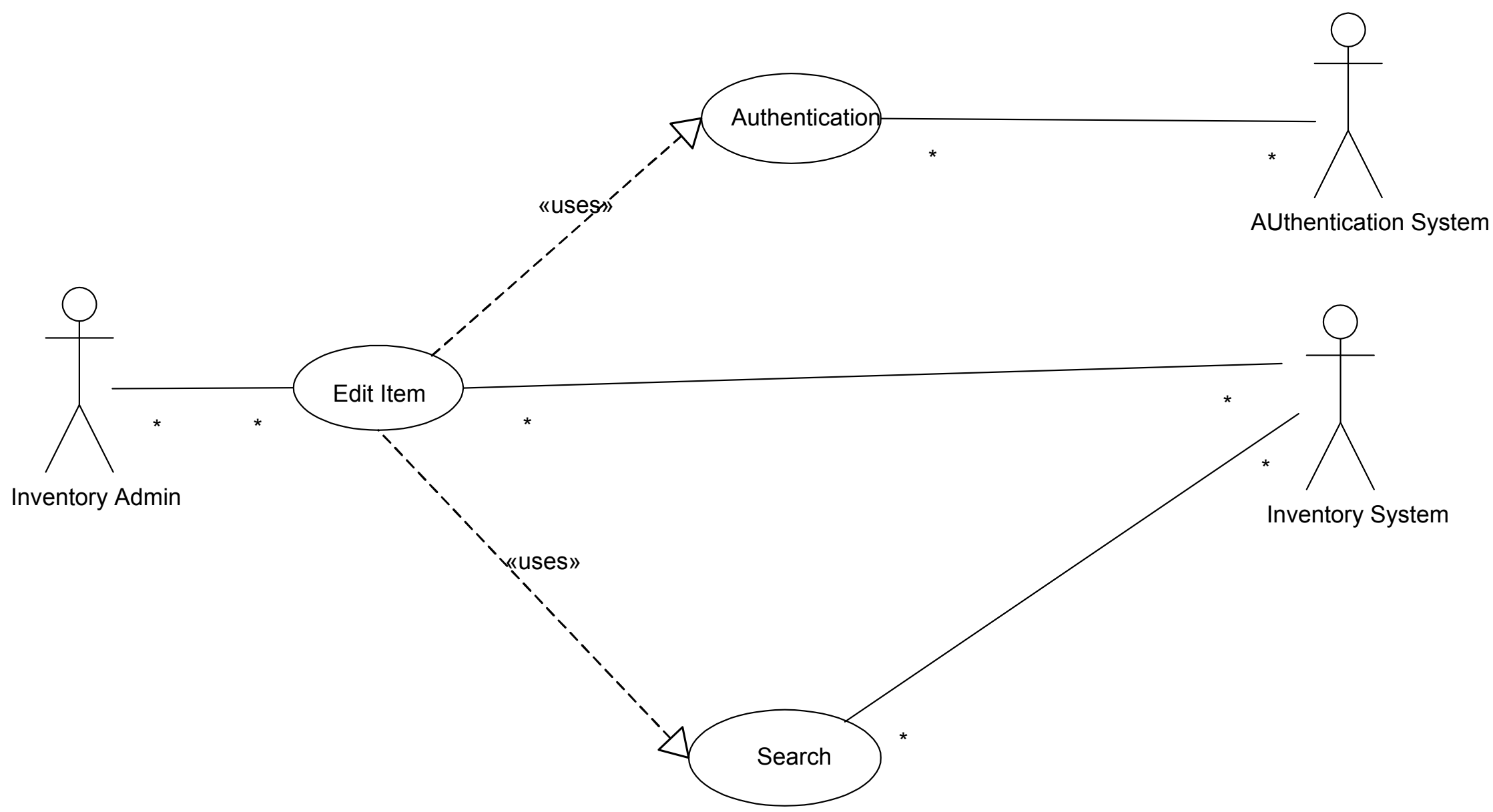

**EDT.UC Edit Use Case**

**Name:** Add New Asset Use Case

**Identifier**: ANI.UC

**Description**

The use case describes the operation of adding a new asset to the inventory.

**Goal**

The Inventory Admin initiates the use case. The use case presents how Inventory Admin can add new inventory asset.

**Preconditions**

1. The Inventory Admin is authenticated

**Assumptions**

1. We assume that use Knows the results of each operation there is no go back actions

**Basic Course**

1. Use case begins when Administrator start a new asset
2. Administrator select asset type
3. Administrator select asset location
4. Administrator select owner
5. Administrator fill all asset properties

**Alternate Course A:**

**None**

**Exceptional Course:**

1. Use case begins when Administrator start a new asset
2. Type does not exist in the list
3. Administrator send exception request to IT to add the new type and the common properties

**Post conditions**

1. Data base is updated

**Actors**
Inventory Admin, Inventory system, Authentication system
**Included Use Cases**

1. Authentication use case

**Notes**
*We suppose that Hosting system and servers support all the operations*

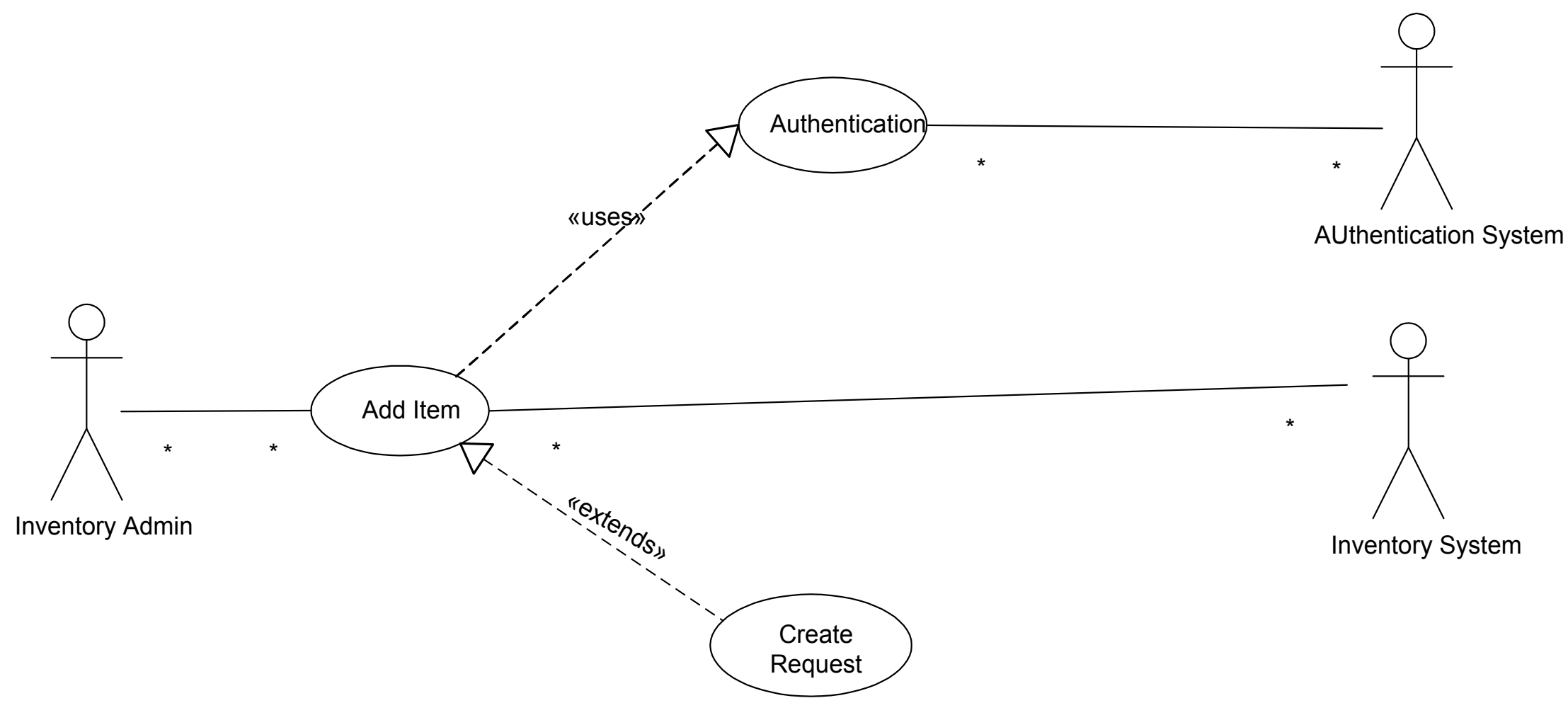

**ANI.UC: Add New Asset Use Case**

**Name:** Create request Use Case
**Identifier**: CRQ.UC
**Description**
The use case describes the activity of creating a new request that the User can do.
**Goal**
The User initiates the use case. The use case presents request process that can be done by User.
**Preconditions**

1. The User is authenticated

**Assumptions**

1. We assume that use Knows the results of each operation
2. Only basic request form is available for level 0
3. Requests forms provide search facilities to users levels 1, 2 and 3

**Basic Course**

1. User select basic request form
2. User type a small request text specifying asset (s) location,..
3. User click on submit

**Alternate Course A:**

1. User select advanced request form
2. User enter asset serial number and location
3. User enter small text describing operation to be done
4. User click on submit

**Alternate Course B:**

1. User select advanced request form
2. User enter asset serial number and location
3. User enter small text describing operation to be done
4. User can add as many assets as he wants to the list by clicking add another asset button
5. User click on submit

**Alternate Course C:**

1. User select exception request form
2. User enter a text message describing the exception that occurs
3. User click on submit

**Exceptional Course :**

1. User search for asset
2. Asset not available to be borrowed
3. Error Message is displayed

**Post conditions**
A new request is pending waiting approval
**Actors**
Inventory Admin, Inventory system, Authentication system

**Included Use Cases**
1. Search use case
2. Authentication use case

**Notes**
*We suppose that Hosting system and servers support all the operations*

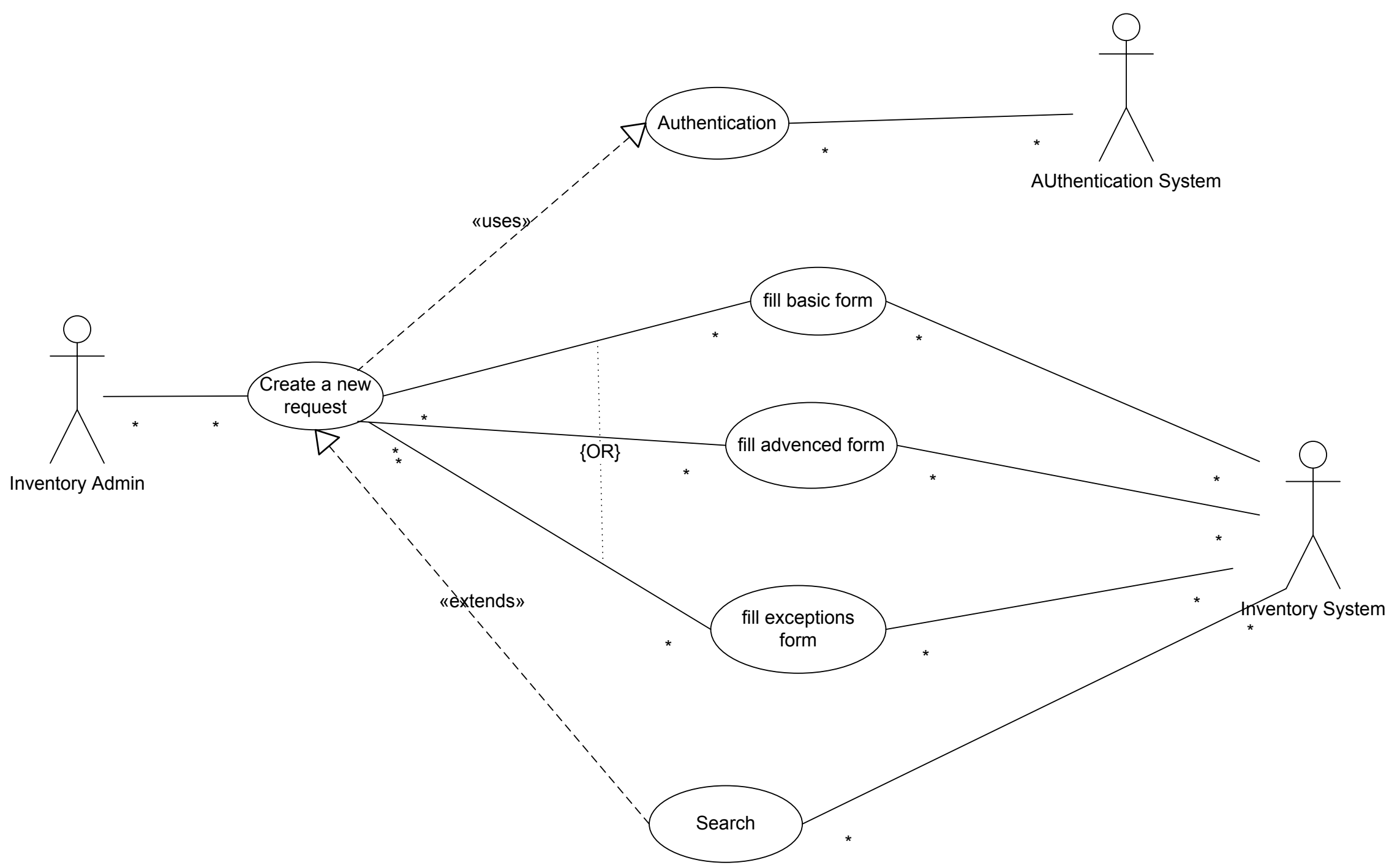

**CRQ.UC: Create request Use Case**

**Name:** Returning Asset
**Identifier**: RTI.UC
**Description**
The use case describes the returning asset update that the Inventory Admin can do.
**Goal**
The Inventory Admin initiates the use case. The use case presents all the updates to the inventory that can be done by the Inventory Admin.
**Preconditions**
1. The Inventory Admin is authenticated

**Assumptions**
1. We assume that use Knows the results of each operation there is no go back actions

**Basic Course**
1. Use case begins when Inventory Admin start returning
2. Inventory Admin select asset state to available

**Alternate Course A:**
**Condition:**

1. Use case begins when Inventory Admin start returning
2. Inventory Admin select asset state to damaged

**Exceptional Course :**

**Post conditions**
1. The inventory system is updated

**Actors**
Inventory Admin, Inventory system, Authentication system

**Included Use Cases**

1. Authentication use case

**Notes**

*We suppose that Hosting system and servers support all the operations*

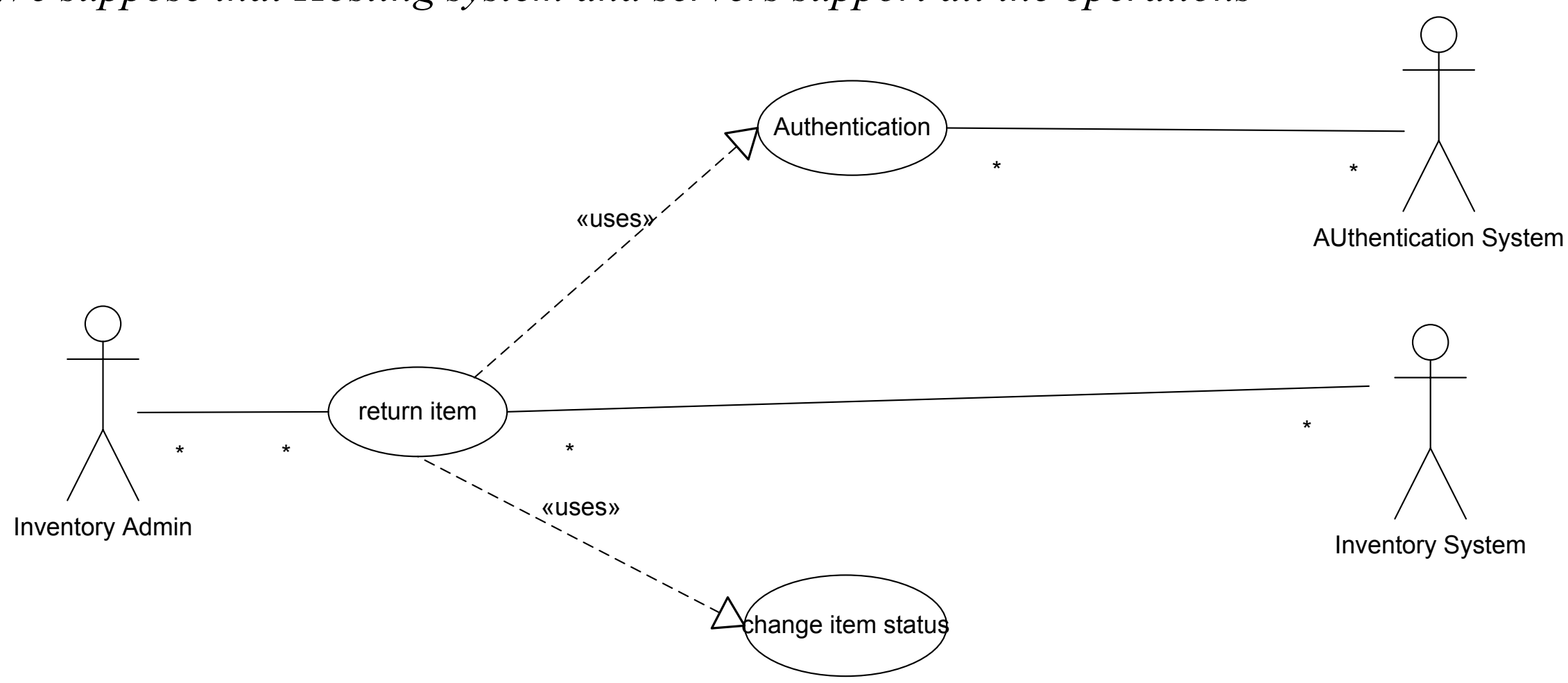

**RTI.UC: Returning Asset**

**Name:** Approving Request
**Identifier**: APR.UC
**Description**
The use case describes the approving of requests that an Inventory Admin can do.
**Goal**
The Inventory Admin initiates the use case. The use case presents all the approval or denial of a user request that can be done by an Inventory Admin.
**Preconditions**
1. The Inventory Admin is authenticated
2. Request list not empty

**Assumptions**
1. We assume that use Knows the results of each operation

**Basic Course**
1. Use case begins when Inventory Admin display pending request list
2. System display only requests that he has privileges to approve
3. Inventory Admin select request that he want to approve
4. Approve is confirmed
5. System send notes to users

**Alternate Course A:**
**Condition:**
1. Use case begins when Inventory Admin display pending request list
2. System display only requests that he has privileges to approve
3. Inventory Admin select request that he want to reject
4. rejection is confirmed
5. System send notes to users

**Exceptional Course :**

**Post conditions**
2. The Inventory system is updated

**Actors**
Inventory Admin, Inventory system, Authentication system

**Included Use Cases**

3. Search use case
4. Edit use case
5. Authentication use case

**Notes**

*We suppose that Hosting system and servers support all the operations*

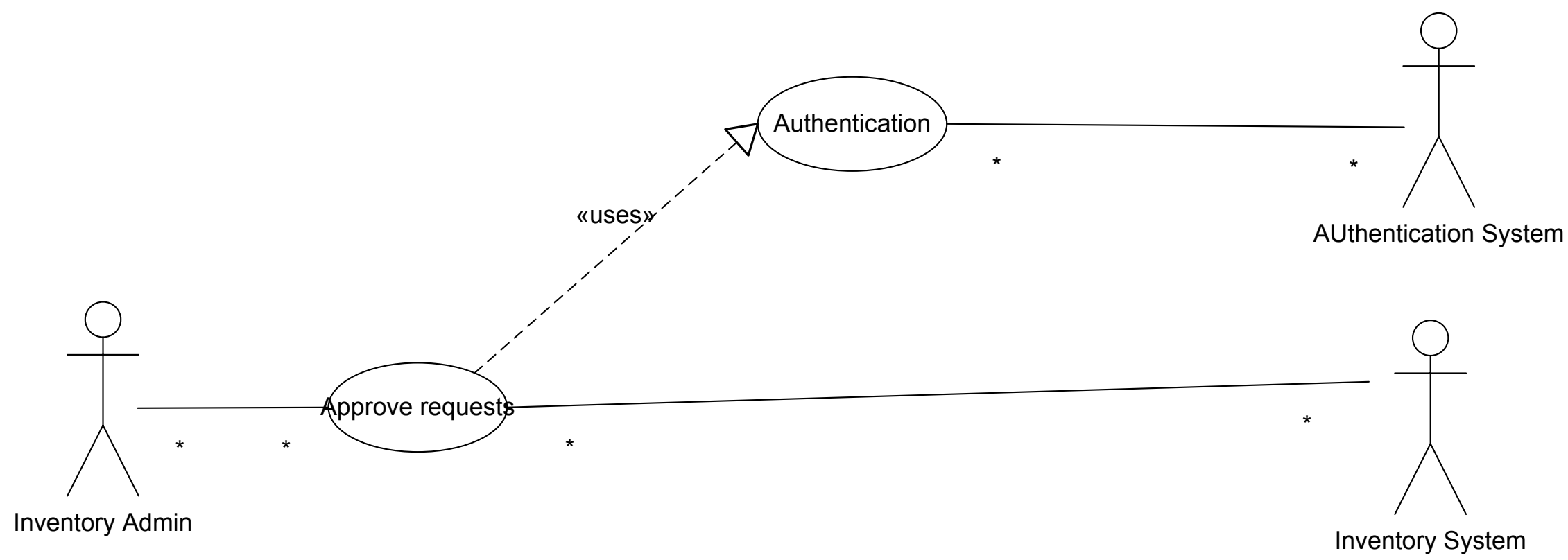

**APR.UC: Approving Request**

**Name:** change permissions Use Case
**Identifier**: CHP.UC
**Description**
The use case describes the modification that Department, Facutly or University Admininistrator can do to a user permissions.
**Goal**
The administrator initiates the use case. The use case presents all change that can be done by the Admininistrators.
**Preconditions**
1. The Administrator is authenticated

**Assumptions**
1. We assume that use Knows the results of each operation
2. Administrator knows the role of each permission on the permission list
3. Administrator cannot assigne permissions more than he have

**Basic Course**
1. Use case begins when Administrator press change permission
2. Administrator select user
3. Administrator modify permissions

**Alternate Course A:**
**Condition:** None

**Exceptional Course :**
1.
   1. Administrator select user
   2. Administrator modify permissions he give permissions more than he have
   3. An error message is displayed

**Post conditions**
2. The permissions of user are changed

**Actors**
Administrator, Inventory system, Authentication system

**Included Use Cases**

1. Authentication use case

**Notes**

*We suppose that Hosting system and servers support all the operations*

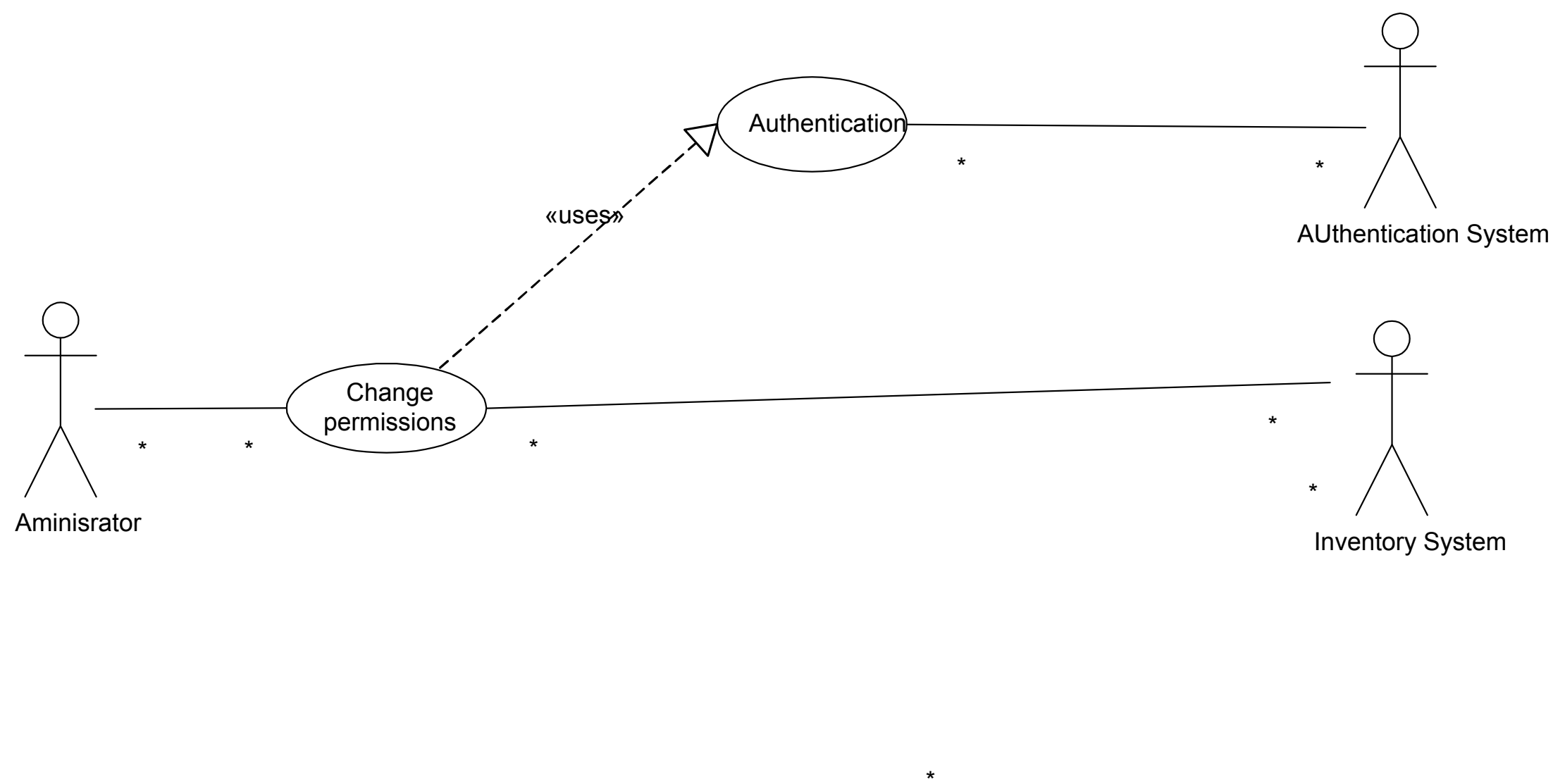

**CHP.UC Change Permissions Use Case**

**Name: Authentication Use Case**
**Identifier**: ATH.UC
**Description**
The use case describes the authentication.
**Goal**
The User initiates the use case. The use case presents authentication operation
**Preconditions**
**None**
**Assumptions**
1. We assume that use Knows the results of each operation

**Basic Course**
1. Use case begins user starts the application
2. User write user name and password
3. User press login
4. User is authenticated
5. User log on to application main page

**Alternate Course A:**
1. None

**Exceptional Course :**
1.
   1. Use case begins user starts the application
   2. User write user name and password
   3. User press login
   4. Authentication fails
   5. An error message is displayed

**Post conditions**
User is authenticated
**Actors**
User, Authentication system

**Included Use Cases**

None

**Notes**

*We suppose that Hosting system and servers support all the operations*

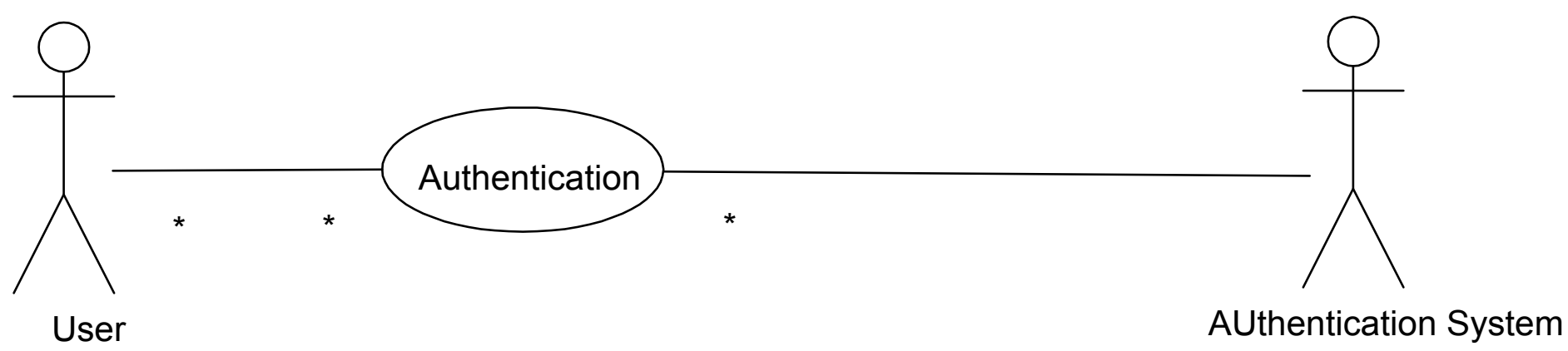

**AUT.UC Authentication Use Case**

**Name:** Search
**Identifier**: SRCH.UC
**Description**
The use case describes the search operation that only authorised users can do.
**Goal**
The User initiates the use case. The use case presents search and advanced search that can be done by the User.
**Preconditions**

1. The User is authenticated
2. Search operation is authorised

**Assumptions**

1. We assume that use Knows the results of each operation

**Basic Course**

1. Use case begins when User start fill searching field
2. Press submit
3. Search results are displayed

**Alternate Course A:**

1. Use case begins when User press advanced search
2. Fills fields
3. Press submit
4. Search results are displayed

**Exceptional Course :**

1. Use case begins when User start fill searching field
2. Press submit
3. No results are found message is displayed

**Post conditions**
**None**

**Actors**
User, Inventory system, Authentication system

**Included Use Cases**

1. Authentication use case

**Notes**

*We suppose that Hosting system and servers support all the operations*

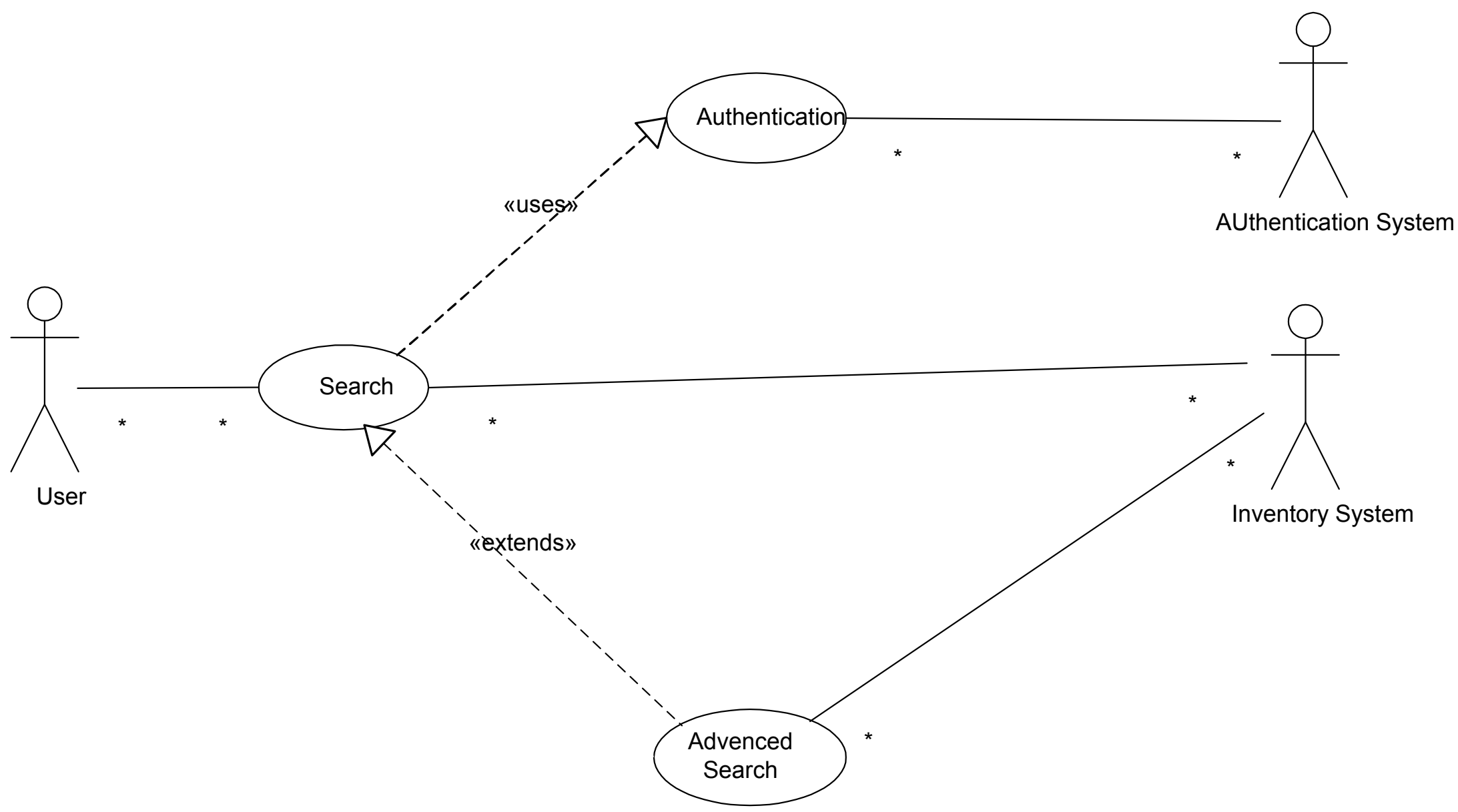

**SRCH.UC  Search Use Case**

**Name:** Create Reports Use Case
**Identifier**: REP.UC
**Description**
The use case describes the creation of reports that the User can do.
**Goal**
The User initiates the use case. The use case presents reports that can be created by the User. Three reports can be created Reports

- User Permission Report
- Request Report
- Assets By Location Report

**Preconditions**
1. The User is authenticated
2. User is authorised to create reports

**Assumptions**
1. We assume that use Knows the results of each operation

**Basic Course**
1. Use case begins when User click on a report type
2. General report is displayed
3. Report can be filtered
4. Fields can be sorted

**Alternate Course A:**
**None**

**Exceptional Course :**
None

**Post conditions**
Report is generated

**Actors**
User, Inventory system, Authentication system

**Included Use Cases**

1. Authentication use case

**Notes**

*We suppose that Hosting system and servers support all the operations*

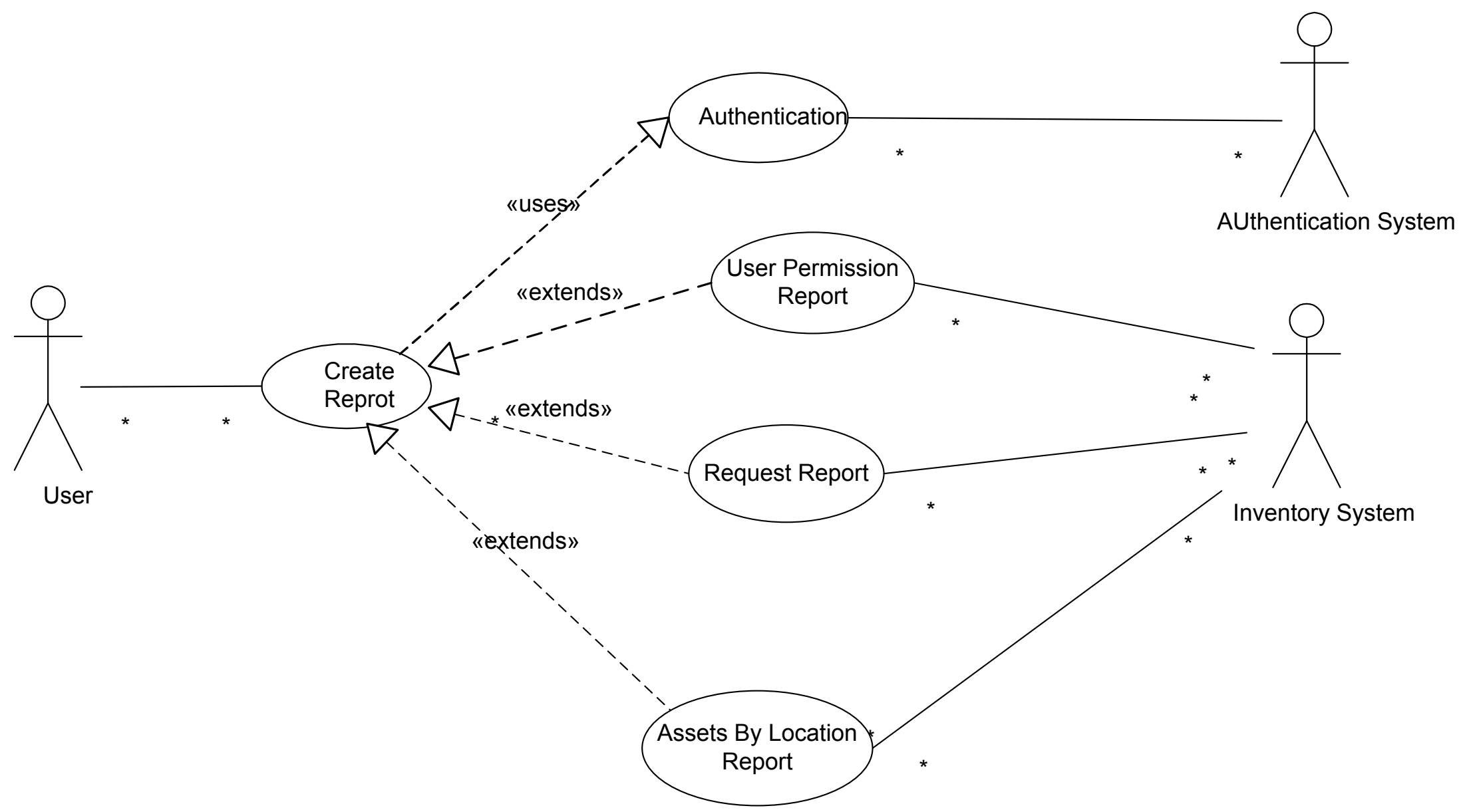

**REP.UC Create reports Use Case**

# 8. Entity relationship diagram

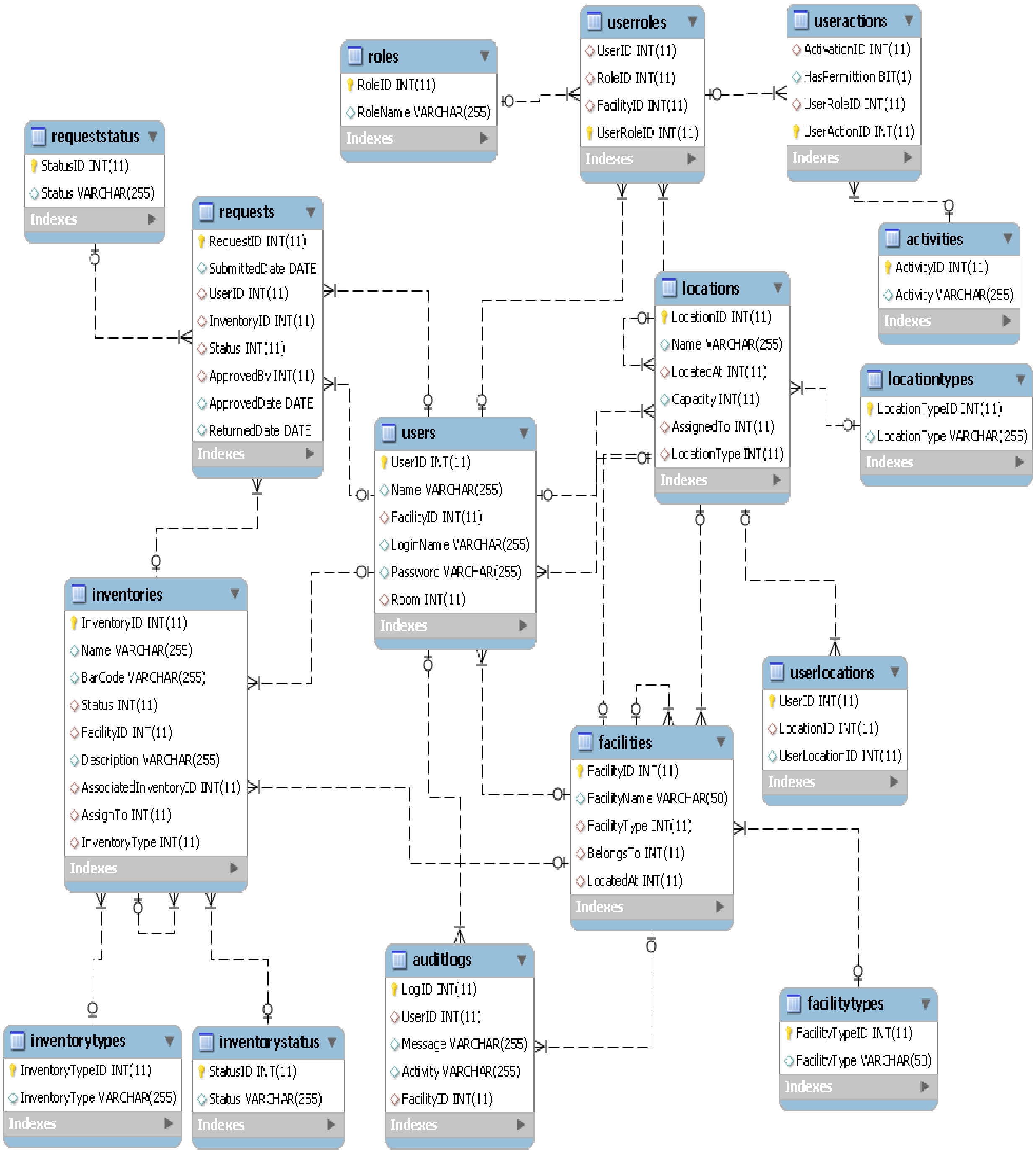

# 9. Cost Estimation (COCOMO)

***Effort Adjustment Factor***

| | Ratings | | | | | |
|---|---|---|---|---|---|---|
| Cost Drivers | Very Low | Low | Nominal | High | Very High | Extra High |
| **Product attributes** | | | | | | |
| Required software reliability | 0.75 | 0.88 | 1 | 1.15 | 1.4 | |
| Size of application database | | 0.94 | 1 | 1.08 | 1.16 | |
| Complexity of the product | 0.7 | 0.85 | 1 | 1.15 | 1.3 | 1.65 |
| **Hardware attributes** | | | | | | |
| Run-time performance constraints | | | 1 | 1.11 | 1.3 | 1.66 |
| Memory constraints | | | 1 | 1.06 | 1.21 | 1.56 |
| Volatility of the virtual machine environment | | 0.87 | 1 | 1.15 | 1.3 | |
| Required turnabout time | | 0.87 | 1 | 1.07 | 1.15 | |
| **Personnel attributes** | | | | | | |
| Analyst capability | 1.46 | 1.19 | 1 | 0.86 | 0.71 | |
| Applications experience | 1.29 | 1.13 | 1 | 0.91 | 0.82 | |
| Software engineer capability | 1.42 | 1.17 | 1 | 0.86 | 0.7 | |
| Virtual machine experience | 1.21 | 1.1 | 1 | 0.9 | | |
| Programming language experience | 1.14 | 1.07 | 1 | 0.95 | | |
| **Project attributes** | | | | | | |
| Application of software engineering methods | 1.24 | 1.1 | 1 | 0.91 | 0.82 | |
| Use of software tools | 1.24 | 1.1 | 1 | 0.91 | 0.83 | |
| Required development schedule | 1.23 | 1.08 | 1 | 1.04 | 1.1 | |

Considering a 3 months project (14w), 8 people working 10 hours/weak =>
1120 hours (/160h) = 7 person months
PM = ai*EAF*KSLOC^bi
KSLOC = (PM/(ai*EAF))^(1/bi)
KSLOC = **1.078**
Pricing per hour = $20
**Project Total = $22,400.00**